\documentclass[notoc]{JHEP3}
\usepackage{amssymb,amsfonts}
\usepackage{amsmath}

\def\d{\partial}
\def\F{{\cal F}}
\def\FF{{\bar\F}}
\def\eps{\varepsilon}
\def\R{\mathbb{R}}
\def\Z{\mathbb{Z}}
\def\C{C_{(0)}}
\def\g{\hat{g}}

\author{N. Couchoud \\
Laboratoire de Physique Th\'eorique et Hautes \'Energies%
\thanks{Unit\'e mixte du CNRS et des Universit\'es de Paris VI et Paris VII,
UMR 7589.} \\
Universit\'e Pierre et Marie Curie,  Paris VI \\
4 place Jussieu, 75252 Paris CEDEX 05, France \\
\email{couchoud@lpthe.jussieu.fr}\\
\\
Laboratoire de Physique Th\'eorique de l'\'Ecole Normale Sup\'erieure%
\thanks{Unit\'e mixte du CNRS et de l'\'Ecole Normale Sup\'erieure,
UMR 8549.} \\
24 rue Lhomond, 75231 Paris CEDEX 05, France \\
\email{couchoud@lpt.ens.fr}
}

\abstract{
We review some facts about $AdS_2\times S^2$ branes in $AdS_3\times
S^3$ with a Neveu-Schwarz background, and consider the case of Ramond-Ramond
backgrounds. We compute the spectrum of quadratic fluctuations in the
low-energy approximation and discuss the open-string geometry.
}
\title{Anti-de Sitter branes with Neveu-Schwarz and Ramond-Ramond backgrounds}
\preprint{ LPTHE-?? \\ LPTENS-03/03 \\ \hepth{0301195} }

\begin{document}
\section{Introduction}
It was argued in \cite{KR} that a ``compactification'' \`a la
Randall-Sundrum can be implemented in string theory by D5-branes in
the near-horizon geometry of D3-branes. One thus obtains generically
an $AdS_4\times S^2$ brane in $AdS_5\times S^5$. As $AdS_5\times S^5$
has a nonzero Ramond-Ramond (RR) background field, one is led to
consider, more generally, $AdS\times S$ geometries with nonzero RR
fluxes.

A simple case is the $AdS_2\times S^2$ brane in $AdS_3\times S^3$,
because the setup with RR backgrounds is S-dual to the well-known
setup with Neveu-Schwarz backgrounds, so we will consider that case
here. After computing the SL$(2,\R)\times$SU(2) content of the
spectrum, we study the open string geometry of the brane. One reason
to study it is the fact, first noticed in \cite{BP}, and conjectured
to be general at least in the probe brane limit, that the
anti-de-Sitter and sphere radii are the same, which, in the case of
the Karch-Randall setup \cite{KR}, would make difficult the
construction of a realistic model from it. So, we compute the
effective radii of the D-brane. We find that:
\begin{itemize}
\item The SL$(2,\R)\times$SU(2) content of the spectrum is the same in
the NS and RR cases, which is expected from S-duality.
\item As in the NS case, the two radii are equal.
\item Contrary to the NS case, where the common radius is equal to the
  radius of the bulk $AdS_3\times S^3$, the common radius in the RR
  case can take any value.
\end{itemize}

The fact that the radii are equal can be shown to be a consequence of
supersymmetry (cf \cite{BP}). More precisely, this seems to be linked
with extended supersymmetry, where the R-symmetry group is the
isometry group of a sphere. So one can hope to obtain realistic brane
worlds from Randall-Sundrum only from $\mathcal{N}\leq1$
supersymmetry.

This paper is organized as follows. In section 2, we recall some facts
about the Neveu-Schwarz case. The novel results of this article are in
section 3. For the reader's convenience, we include the proof of the
SL$(2,\R)$ symmetry of the low-energy action of the D3-brane in the
appendix, with the transformations of the fields.

\boldmath
\section{A reminder of $AdS_2\times S^2$ D-branes in $AdS_3\times S^3$
with pure Neveu-Schwarz background}
\unboldmath
\subsection{Notations}
We parametrize $AdS_3\times S^3$ with coordinates such that the metric
and the NS two-form read:
\begin{equation}
\begin{array}{c}
ds^2 = L^2\,[d\psi^2+\cosh^2\psi\,(d\omega^2-\cosh^2\omega\,d\tau^2)
           +d\chi^2+\sin^2\chi\,(d\theta^2+\sin^2\theta\,d\phi^2)] \\ \\
\displaystyle
B=L^2\left[
  \left(\psi+\frac{\sinh2\psi}{2}\right)\cosh\omega\,d\omega\wedge d\tau +
  \left(\chi-\frac{\sin2\chi}{2}\right)\sin\theta\,d\theta\wedge d\phi
\right]
\end{array}
\end{equation}
and the dilaton is a constant $\Phi=\Phi_{NS}$.

Symmetric D-branes in this geometry are hypersurfaces with
$\psi=\psi_0$ and $\chi=\chi_0$, so their geometry is $AdS_2\times
S^2$ with induced radii $L\cosh\psi_0$ and $L\sin\chi_0$. They can be
stabilized by an appropriate electromagnetic field:
\begin{equation}
F_0 = -\frac{L^2}{2\pi\alpha'}\left(\psi_0\cosh\omega\,d\omega\wedge d\tau
       + \chi_0\sin\theta\,d\theta\wedge d\phi\right).
\end{equation}

Their low-energy effective action is given by the Dirac-Born-Infeld
plus Wess-Zumino (DBI-WZ) action, which in the most general case reads
\begin{equation}
\label{action}
S = T_{D3}\left[ -\int d^4x e^{-\Phi}\sqrt{-\det(\g+\F)} 
    + \int\left(\hat{C}_{(4)} + \hat{C}_{(2)}\wedge\F
                + \frac{1}{2}\C\F\wedge\F\right)
\right]\,,
\end{equation}
where $\F = 2\pi\alpha'F + \hat{B}$. (As $C_{(4)}$ will be zero in
this whole article, it will not be mentioned again.)

\subsection{Quantized charges}
There are two quantized charges:
\begin{itemize}
\item 
$\displaystyle p=-\frac{1}{2\pi}\int_{S^2} d\theta\,d\phi\,F_{\theta\phi}$
is an integer because of Dirac quantization. It can be interpreted as
the D-string charge of the D3-brane.
\item The F-string charge should, of course, also be an integer; since
F-strings couple with the NS two-form, this charge reads
\begin{equation}
q = \frac{1}{T_F} 
    \int_{S^2} d\theta\,d\phi\,\frac{\delta S}{\delta B_{\omega\tau}} 
\equiv \frac{1}{2\pi} \int_{S^2} d\theta\,d\phi\,{\tilde F}_{\theta\phi}\,;
\end{equation}
the explicit form of $\tilde F$ is given in the appendix.
\end{itemize}
Notice that these facts are for any D3-brane with topology $M\times S^2$.

In our case, one has $\displaystyle p=\frac{L^2}{\pi\alpha'}\chi_0$ and
$\displaystyle q=-\frac{L^2}{\pi\alpha'}e^{-\Phi}\sin\chi_0\sinh\psi_0$.%
\footnote{Notice that the quantization condition given in \cite{BP} is
true only in the small $\chi_0$ limit.} The fact that these quantities
are quantized means two things:
\begin{itemize}
\item The set of possible positions for the branes is discrete; in
particular, the $\chi_0$ parameter can have only a finite number of
values.
\item The fluctuations of the branes must be compatible with these
quantization conditions; this implies the stability of the branes (see
\cite{BDS}) and some restrictions on the spectrum.
\end{itemize}
It also implies a restriction of the SL$(2,\R)$ symmetry group of the
DBI-WZ action to the S-duality group SL$(2,\Z)$, which is precisely
the subgroup of SL$(2,\R)$ which preserves the quantization. This
SL$(2,\Z)$ is conjectured to be an exact symmetry of the full string
theory.

\subsection{Quadratic fluctuations}
\label{fluc}
The spectrum of quadratic fluctuations was computed in \cite{PR}.
They are obtained by taking
\begin{equation}
\psi=\psi_0+\delta\psi\,,\ \chi=\chi_0+\delta\chi\,,\text{ and }
F=F_0+\frac{L^2}{2\pi\alpha'}f \text{ with } df=0,
\end{equation}
and expanding the action up to quadratic terms in $\delta\psi$,
$\delta\chi$ and $f$. The expansion begins with two linear terms that
are proportional to $\int d\theta\,d\phi\,f_{\theta\phi}$, which is
zero because of the quantization condition, and 
$\int d\omega\,d\tau\,f_{\omega\tau}$, which is zero if the electromagnetic
potential behaves well at infinity because $AdS_2$ is topologically
trivial.

The quadratic terms read:
\begin{equation}
\label{quadNS}
\begin{array}{l}
\displaystyle\delta^{(2)}S = T_{D3}L^4e^{-\Phi}\cosh\psi_0\sin\chi_0
             \int d\omega\,d\tau\,d\theta\,d\phi\sqrt{-\det\gamma}\\
\displaystyle\left[
  -\frac{1}{2}(\d_m\delta\psi\d^m\delta\psi + \d_m\delta\chi\d^m\delta\chi)
  -\frac{1}{4}f_{mn}f^{mn} +(\delta\psi)^2 -(\delta\chi)^2
  -2\delta\psi\frac{f_{\omega\tau}}{\cosh\omega}
  +2\delta\chi\frac{f_{\theta\phi}}{\sin\theta} \right] \\
\displaystyle-T_{D3}L^4e^{-\Phi}\sinh\psi_0\cos\chi_0\int f\wedge f
\end{array}
\end{equation}
where $\gamma$ is the $AdS_2\times S^2$ metric with radii 1, and
indices are raised with it.

It is then easy to find the SL$(2,\R)\times$SU(2) content of the
spectrum, which is explicitly given in \cite{PR}, where it is also
shown that it agrees with the results of conformal field theory
(except for the brane-dependent cut-off on the allowed angular
momenta).

\boldmath
\section{$AdS_2\times S^2$ D-branes in $AdS_3\times S^3$
with RR background}
\unboldmath
\subsection{The background}
The RR background is obtained from the NS background by the following
S-duality transformation: $C_{(2)}\to B$; $B\to-C_{(2)}$; $F\to{\tilde F}$
and $\tau\to-1/\tau$, where $\tau=\C+ie^{-\Phi}$ (this corresponds to
$(r,s,t,u)=(0,-1,1,0)$ in equation \eqref{S-transf}). It is shown in
the appendix that ${\tilde F}$ is then transformed into $-F$. Thus,
the quantum numbers $(p,q)$ become $(-q,p)$, which means that D-string
and F-string charge are exchanged.

The fields obtained this way are
\begin{equation}
\begin{array}{rcl}
\Phi & = & -\Phi_{NS} \\
C_{(2)} & = & B_{NS} \\
F_0 & = &
\displaystyle
\frac{e^{\Phi}L_{NS}^2}{2\pi\alpha'}
    (\cos\chi_0\cosh\psi_0\cosh\omega\,d\omega\wedge d\tau
    + \sin\chi_0\sinh\psi_0\sin\theta\,d\theta\wedge d\phi).
\end{array}
\end{equation}

The string metric is also changed by this transformation: since the
Einstein metric $g_E = e^{-\Phi/2}g$ is invariant, the string metric
becomes $ds^2 = e^{\Phi}ds^2_{NS}$. In what follows, we redefine
$L^2 \equiv L_{RR}^2 = e^{\Phi}L_{NS}^2$.

\subsection{Quadratic fluctuations}
Since the S-duality transformation is a symmetry of the equations of
motion deduced from the low-energy action, it is expected that the
SL$(2,\R)\times$SU(2) content of the spectrum be the same as in the
previous section. Anyway, it is useful to check explicitely that
S-duality works properly in that case.

To do this, we expand the action \eqref{action} in the fluctuations as
in section \ref{fluc}. As in the NS case, the linear terms disappear.

At the first look, the quadratic terms look messy, since $\delta\psi$,
$\delta\chi$, $f_{\theta\phi}$ and $f_{\omega\tau}$ are coupled in all
possible ways, contrary to the NS case. However, by taking
\begin{equation}
\begin{array}{l}
\delta_A=\sinh\psi_0\cos\chi_0\,\delta\psi-\cosh\psi_0\sin\chi_0\,\delta\chi
\\
\delta_S=\cosh\psi_0\sin\chi_0\,\delta\psi+\sinh\psi_0\cos\chi_0\,\delta\chi
\end{array}
\end{equation}
the quadratic terms read
\begin{equation}
\label{quadR}
\begin{array}{l}
\displaystyle\delta^{(2)}S = T_{D3}L^4e^{-\Phi}
             \frac{\cosh\psi_0\sin\chi_0}{\sin^2\chi_0+\sinh^2\psi_0}
             \int d\omega\,d\tau\,d\theta\,d\phi\sqrt{-\det\gamma}\\
\displaystyle\left[
  -\frac{1}{2}(\d_m\delta_A\d^m\delta_A + \d_m\delta_S\d^m\delta_S)
  -\frac{1}{4}f_{mn}f^{mn} +(\delta_A)^2 -(\delta_S)^2
  -2\delta_A\frac{f_{\omega\tau}}{\cosh\omega}
  +2\delta_S\frac{f_{\theta\phi}}{\sin\theta} \right] \\
\displaystyle+T_{D3}L^4e^{-\Phi}
  \frac{\sinh\psi_0\cos\chi_0}{\sin^2\chi_0+\sinh^2\psi_0}\int f\wedge f
\ .
\end{array}
\end{equation}
The similarity with equation \eqref{quadNS} implies that the
SL$(2,\R)\times$SU(2) content of the spectrum is the same, as
expected.

As is well-known, in the NS case, the exact spectrum, as determined
from CFT, has a brane-dependent cut-off: the maximal allowed spin is
half of the magnetic charge $p$. If S-duality is true, we expect that
in the RR case the maximal allowed spin is half of the electric
charge. Such a result could not be found directly, since we have no
sigma-model for the world-sheet of strings with a nonzero RR
background.

\subsection{The effective metric}
As is well-known, for any D-brane without RR backgrounds, the
fields on the brane can be considered as coupled to an effective
metric (open string metric) given by the formula
\begin{equation}
\label{effmet}
g_o^{-1}=[(g+\F)^{-1}]_S
\end{equation}
where the index $S$ means symmetrizing the matrix. In the case of the
$AdS_2\times S^2$ D-brane in $AdS_3\times S^3$, it was noticed in
\cite{BP} that both radii are equal to $L$ independently of the
position of the brane, although the induced radii depend on it and can
be very different of each other.

It is interesting to see whether there is, in the RR case, a notion of
effective metric with such properties. From the fluctuations found in
the previous section, it is clear that the effective geometry is
$AdS_2\times S^2$ with equal radii. The absolute normalization of
these radii cannot be deduced from the previous calculations only.

A way to find it is to T-dualize our setup along a dimension
orthogonal to $AdS_3\times S^3$ to obtain an $AdS_2\times S^2\times
S^1$ D-brane in $AdS_3\times S^3\times S^1$. The nonzero resulting
background fields are the following (with $x$ the additional dimension):
\begin{equation}
\begin{array}{rcl}
ds^2 & = & ds^2(\mathrm{AdS}_2\times\mathrm{S}^2) + dx^2 \\
C_{mnx} & = & C_{mn}(\mathrm{AdS}_2\times\mathrm{S}^2)
\end{array}
\end{equation}
the dilaton is, as previously, a constant, and the electromagnetic
field is the same. It is then easy to see that the WZ term of the
action is left unchanged (except for the fact that $T_{D3}$ is
replaced by $T_{D4}$), so that all additional terms come from the
determinant in the DBI term. 

After some straightforward calculation, one finds that the quadratic
terms of the action are essentially the same as in eq. \eqref{quadR},
except for the following facts:
\begin{itemize}
\item $T_{D3}$ is replaced by $T_{D4}$.
\item $f\wedge f$ is replaced by $f\wedge f\wedge dx$.
\item More importantly, new terms are added inside the bracket:
\begin{equation}
L^2(\sin^2\chi_0+\sinh^2\psi_0)
\left[ -\frac{1}{2}(\d_x\delta_A\d^x\delta_A + \d_x\delta_S\d^x\delta_S)
  -\frac{1}{2}f_{mx}f^{mx} \right]\ .
\end{equation}
\end{itemize}
From these new terms, one immediately obtains that the absolute
normalization of the $AdS_2\times S^2$ effective radius is
\begin{equation}
R^2=L^2(\sin^2\chi_0+\sinh^2\psi_0).
\end{equation}
It is easy to check that this is what one finds from the formula
\eqref{effmet} without any correction.

So, the situation is different from the NS case, since the radius
depends on the position of the brane, and can take any value, which
seems to contradict S-duality. However, as S-duality is a symmetry of
the equations of motion, and not of the action, this is not
surprising.

In particular, in the limit $\psi_0 \to \infty$, one obtains an
infinite radius, {\em i.e.} flat space. The question whether this
could lead to a realistic brane universe would certainly be
interesting to investigate.

\acknowledgments
I thank C.~Bachas for long discussions and carefully reading the
manuscript. I also thank S.~Ribault and P.~Windey for useful
discussions.

\appendix
\boldmath
\section{SL$(2,\R)$ symmetry of the D3-brane}
\unboldmath
We include here, for the reader's convenience, the proof of the
SL$(2,\R)$ symmetry of the low-energy effective action of the
D3-brane. We essentially follow the lines of \cite{SD1,SD2,SD3}.

For the antisymmetric symbol, we use the convention
$\eps_{0123} = \eps^{0123} = +1$. (With this convention, $\eps$ is
{\em not} a tensor.)

The Born-Infeld and Wess-Zumino terms of the action read:
\begin{equation}
S = T_{D3}\left( -\int d^4x \sqrt{-\det(\g_E+e^{-\Phi/2}\F)} 
    + \int\hat{C}_{(2)}\wedge\F + \frac{1}{2}\int \C\F\wedge\F \right)\,,
\end{equation}
where $\F = 2\pi\alpha'F + \hat{B}$ and $g_E=e^{-\Phi/2}g$ is the
Einstein metric. The equation of motion can be written $d{\tilde F}=0$,
where
\begin{equation}
2\pi\alpha'{\tilde F}_{mn} = 
  \eps_{mnpq}\,e^{-\Phi/2}\,\frac{\d L_{BI}}{\d\FF_{pq}}
   + \hat{C}_{mn} + \C\F_{mn}\,,
\end{equation}
\begin{equation}
\frac{\d L_{BI}}{\d\FF_{pq}} =
   \frac{1}{2} \sqrt{-\det(\g_E+\FF)}\,[(\g_E+\FF)^{-1}]^{[pq]}
\text{ with } \FF = e^{-\Phi/2}\F.
\end{equation}

We want to show that there is an SL$(2,\R)$ symmetry acting as:
\begin{equation}
\label{S-transf}
\begin{array}{c}
\left(\begin{array}{l} C_{(2)} \\ B \end{array}\right) \to
\left(\begin{array}{cc} r & -s \\ -t & u \end{array}\right)
\left(\begin{array}{l} C_{(2)} \\ B \end{array}\right)\;;\;
\left(\begin{array}{l} \tilde{F} \\ F \end{array}\right) \to
\left(\begin{array}{cc} r & s \\ t & u \end{array}\right)
\left(\begin{array}{l} \tilde{F} \\ F \end{array}\right)\;;
\\ \\
\displaystyle \tau \to \frac{r\tau+s}{t\tau+u}\,,
\text{ where } \tau = \C + ie^{-\Phi}.
\end{array}
\end{equation}
To do this, we consider the infinitesimal transformations
\begin{equation}
\begin{array}{c}
\delta \C = 2a\C + b - c(\C^2-e^{-2\Phi}) \\
\delta(e^{-\Phi/2}) = e^{-\Phi/2}(a-c\C) \\
\delta\F = -a\F + c{\tilde\F}
  \text{ with } {\tilde\F}=2\pi\alpha'{\tilde F}-\hat{C}_{(2)}
\end{array}
\end{equation}
and we want to show that
\begin{equation}
\delta{\tilde\F} = a{\tilde\F} + b\F.
\end{equation}

The $b$ term is trivial.

Noticing that
\begin{equation}
\delta\FF = ce^{-\Phi/2}({\tilde \F} - \C\F)
\end{equation}
it is easy to show that the $a$ term is what we expect.

After a straightforward calculation, the $c$ term reads
\begin{equation}
\delta_c {\tilde \F}_{mn} = c\left(
  e^{-3\Phi/2}\eps_{abcd}\,\frac{\d L_{BI}}{\d\FF_{cd}}\,
        \eps_{mnpq}\,\frac{\d^2L_{BI}}{\d\FF_{ab}\d\FF_{pq}}
  + e^{-2\Phi}\F_{mn}\right).
\end{equation}
Using the commutativity of the partial derivatives and the symmetry
properties of the $\eps$ symbol, we obtain
\begin{equation}
\delta_c {\tilde \F}_{mn} = c\left(
  \frac{1}{2}\,e^{-3\Phi/2}\eps_{mnpq} \frac{\d}{\d\FF_{pq}}\left(
    \eps_{abcd}\,\frac{\d L_{BI}}{\d\FF_{ab}}\,\frac{\d L_{BI}}{\d\FF_{cd}}
  \right) + e^{-2\Phi}\F_{mn}\right).
\end{equation}
The antisymmetric part of $(\g_E+\FF)^{-1}$ reads
\begin{equation}
[(\g_E+\FF)^{-1}]^{[pq]} =
  [(\g_E+\FF)^{-1}]^{pp'} \FF_{p'q'} [(\g_E+\FF)^{-1}]^{qq'}
\end{equation}
and one has, for any matrix $M$,
\begin{equation}
\eps_{abcd}M^{aa'}M^{bb'}M^{cc'}M^{dd'} = \eps^{a'b'c'd'} \det M
\end{equation}
so that
\begin{equation}
\eps_{abcd}\,\frac{\d L_{BI}}{\d\FF_{ab}}\,\frac{\d L_{BI}}{\d\FF_{cd}} =
-\frac{1}{4}\eps^{abcd}\FF_{ab}\FF_{cd}
\end{equation}
and finally $\delta_c{\tilde\F}=0$, which is what we want.

\end{document}